\documentclass[prl,twocolumn,floatfix,showpacs]{revtex4}
\usepackage{dcolumn,bm,amsmath}
\usepackage[dvips]{graphicx}
\def\veps{\varepsilon}

\newcommand{\eref}[1]{(\ref{#1})}
\newcommand{\Eref}[1]{Eq.~(\ref{#1})}
\newcommand{\tref}[1]{Table~\ref{#1}}

\begin{document}
\title{Enhancement of the electric dipole moment of the electron
in PbO}
\author{ M. G. Kozlov}
\email{mgk@MF1309.spb.edu} \affiliation{Petersburg Nuclear Physics
Institute, Gatchina, 188300, Russia}
\author{ D. DeMille}
\email{david.demille@yale.edu} \affiliation{Physics Department,
Yale University, New Haven, CT 06520}
\date{\today}

\begin{abstract}
The $a(1)$ state of PbO can be used to measure the electric dipole
moment of the electron $d_e$.  We discuss a semiempirical model
for this state, which yields an estimate of the effective electric
field on the valence electrons in PbO. Our final result is an
upper limit on the measurable energy shift, which is significantly
larger than was anticipated earlier: $ 2|W_d|d_e \ge 2.4\times
10^{25} \textrm{Hz} \left[ \frac{d_e}{e\,\textrm{cm}} \right]$.
\end{abstract}

\pacs{32.80Ys, 11.30.Er, 31.30.Jv}

\maketitle


In his pioneering work, Sandars pointed out that the effective
electric field on a valence electron in a heavy atom is enhanced
by a factor $\sim \! \alpha^2 Z^3$ relative to the applied
laboratory field \cite{San65}. That started a long search for the
electric dipole moment (EDM) of the electron $d_e$ in atomic
experiments (\cite{Com99}). The most stringent limit on $d_e$
follows from an experiment on atomic Tl (Z=81) \cite{RCSD02}.

Even larger enhancement is present in heavy polar diatomic
molecules~\cite{SF78, KL95,Com99}. The heavy atom there is
subjected to an internal E-field of $\sim$1~a.u. $\approx 5\cdot
10^9$~V/cm, which is further enhanced by the relativistic factor
$\alpha^2 Z^3$. This effective field is many orders of magnitude
larger than available laboratory fields; this makes diatomic
molecules very attractive systems to look for $d_e$.

Since $d_e$ is linked to the electron spin, one must work either
with radicals, which have an unpaired electron in the ground
state, or with excited states of ``normal'' molecules. Diatomic
radicals with the ground state $\Sigma_{1/2}$ have large
enhancement factors which can be relatively easily calculated
\cite{Koz85,KE94}. The first results of an EDM measurement in such
a molecule (YbF) were recently published \cite{Hin97,HST02}.  The
molecule PbO is a favorable candidate for a search for $d_e$ in
the excited state $a(1)$ \cite{SF78,DBB00}, and the group at Yale
has begun EDM experiments on PbO~\cite{DBB01}. It is therefore
timely to estimate the effective internal field for the state
$a(1)$ of PbO.

The interaction of $d_e$ with an electric field $\bm{E}$ can be
written in four-component Dirac notation as~\cite{Khr91}:
\begin{eqnarray}
    H_d=2d_e
    \left(\begin{array}{cc}
    0 & 0 \\
    0 & \bm{\sigma E} \\
    \end{array}\right).
\label{00}
\end{eqnarray}
After averaging over the electronic wavefunction this interaction
can be expressed in terms of an effective spin-rotational
Hamiltonian~\cite{Koz85,KL95}:
\begin{equation}
  H_d^{\textrm eff} = W_d\, d_e\, (\bm{J}_e \cdot \bm{n}),
\label{01}
\end{equation}
where $\bm{J}_e$ is the electronic angular momentum and $\bm{n}$
is the unit vector along the molecular axis. In this paper we
estimate $W_d$ for the molecular state $a(1)$:
\begin{equation}
   W_d \equiv d_e^{-1} \langle a(1) |H_d|a(1) \rangle,
\label{02}
\end{equation}
where we used that $\langle a(1)|\bm{J}_e \cdot \bm{n}|a(1)\rangle
\equiv \Omega[a(1)]=1$.  The $\Omega$-doubling for states with
$\Omega=1$ is very small and even in a weak external electric
field the energy eigenstates correspond to definite $\Omega$
rather than definite parity. The energy of the molecule can be
then written as:
\begin{equation}
   W(J,M,\Omega) = BJ(J+1)+ W_d\, d_e \Omega -\frac{DE_0\Omega M}{J(J+1)},
\label{03}
\end{equation}
where $B$ is the rotational constant, $E_0$ is the external
electric field, and $D$ is the molecular dipole moment. The EDM
contribution can be determined from the difference:
\begin{equation}
   W(J,M,\Omega)-W(J,\!-M,\!-\Omega)= 2 W_d\, d_e.
\label{04}
\end{equation}

In order to estimate the matrix element in Eq.~\eref{02} we
construct here a semiempirical wavefunction of the state $a(1)$.
We use the MO~LCAO approach, where each molecular orbital is
expressed as a linear combination of atomic orbitals, and all
molecular matrix elements are reduced to the sums of atomic matrix
elements. The HFS or SO interactions as well as the EDM
enhancement factor grow very rapidly with nuclear charge $Z$.
Therefore, we are interested only in the Pb part of the MO~LCAO
expansion.


Analysis of the molecular observables requires knowledge of
several atomic matrix elements for Pb. We calculate these in the
Dirac-Fock approximation both for neutral Pb and for
$\textrm{Pb}^+$. Results are given in \tref{tab1} for the orbitals
$6s$ and $6p_j$. For the HFS operator we calculate the parameters
$h_{k,k'}$ as defined in \cite{KL95} (we use atomic units unless
the opposite is stated explicitly):
\begin{eqnarray}
        h_{k,k'} &=& -\frac{g_n \alpha}{2 m_p}
        \int \limits_{0}^{\infty} (f_{k} g_{k'}+ g_{k} f_{k'}) dr,
\label{a1}
\end{eqnarray}
where $g_n=0.59$ is the nuclear g-factor of $^{207}\textrm{Pb}$,
$m_p$ is the proton mass, $f_k$ and $g_k$ are upper and lower
components of the Dirac orbitals, and $k = (l-j)(2j+1)$ is the
relativistic quantum number.  For the EDM operator \eref{00} in
our minimal basis set there is only one nonzero radial integral,
between $6s_{1/2}$ and $6p_{1/2}$ orbitals:
\begin{equation}
     w_{sp} =
     -\int \limits_0^{\infty} g_{-1} g_1
     \frac{{\rm d}\phi}{{\rm d}r} r^2 dr,
\label{a2}
\end{equation}
where $\phi$ is the atomic electrostatic potential. Finally, we
also need the atomic SO constant $\xi$ for the $6p$ shell:
\begin{equation}
        H_{\rm SO} = \xi {\bm l \cdot s},
        \quad  \Rightarrow \quad
        \xi = \frac{2}{3}\,(\veps_{6p_{3/2}}-\veps_{6p_{1/2}}).
\label{a4}
\end{equation}

\begin{table}
\caption{Atomic parameters $h_{k,k'}$ (GHz), $w_{sp}$ (a.u.), and
$\xi$ ($\textrm{cm}^{-1}$), calculated in the Dirac-Fock
approximation for Pb and Pb$^+$. The relativistic quantum number
$k$ is equal to $-1$, 1, and $-2$ for $s_{1/2}$, $p_{1/2}$, and
$p_{3/2}$ correspondingly.}

\label{tab1}

\begin{tabular}{lccccccc}
\hline \hline
       & $h_{-1,-1}$ & $h_{1,1}$ & $h_{1,-2}$
       & $h_{-2,-2}$ & $w_{sp}$
       & $\frac{w_{sp}^{\vphantom{1}}}{\sqrt{-h_{-1,-1}h_{1,1}}}$
       & $\xi$ \\
\hline Pb$^+$ & 45.5        & $-$8.9    & $-$1.1
       & 1.8         & $-$34.1   & $-$1.7
       & 9452    \\
Pb     & 42.3        & $-$7.5    & $-$0.9
       & 1.4         & $-$30.0   & $-$1.7
       & 8077    \\
ratio  & 1.08        & 1.20      & 1.25
       & 1.30        & 1.14      & 1.0
       & 1.17    \\
\hline \hline
\end{tabular}
\end{table}

The last row of \tref{tab1} gives the ratio of the radial
integrals for the ion and for the atom. Note that the ratios are
similar for all relevant integrals, and a simple relation between
$w_{sp}$ and HFS constants holds for both cases:
\begin{equation}
        w_{sp} = -1.7 \sqrt{-h_{-1,-1}h_{1,1}}.
\label{a5}
\end{equation}
This relation is critical for semiempirical models of the EDM
enhancement: it implies that the value of the EDM
enhancement does not depend on what set of radial integrals
(i.e. atomic or ionic) is used. Similarly \eref{a5} holds for
other principal quantum numbers $n$, \textit{e.g.} for $7s$ and
$7p_j$.  The particular choice of the radial integrals
becomes important for the normalization of the wavefunction. The
Pb atom in PbO is positively charged; therefore, below we use the
ionic set of integrals from Table~\ref{tab1}.


In order to develop a semi-empirical model for the state $a(1)$,
we have found it necessary to also consider the wavefunctions of
several low-lying states of PbO. Previous work has shown that
these states correspond to the configurations and nominal
$\Lambda, \Sigma$-coupling terms as follows \cite{ODZ75,BP83}:
\begin{eqnarray}
X(0^+)[^1\Sigma^+_0]
\!&:&\!
\sigma_1^2 \sigma_2^2 \sigma_3^2\pi_1^4, \label{b1a}
\\
a(1)[^3\Sigma^+_1\!], C(0^+)[^3\Sigma^-_0\!], C'(1)[^3\Sigma^-_1\!]
\!&:&\!
\sigma_1^2 \sigma_2^2 \sigma_3^2 \pi_1^3 \pi_2,
\label{b1b}
\\
A(0^+)[^3\Pi_0], B(1)[^3\Pi_1], D(1)[^1\Pi_1]
\!&:&\!
\sigma_1^2 \sigma_2^2 \sigma_3 \pi_1^4 \pi_2,
\label{b1c}
\end{eqnarray}
In the naive ionic model of PbO, the two $6p$ electrons from Pb
move to O and close its $2p$-shell. This suggests that the
orbitals $\sigma_{1,2}$ and $\pi_1$ are centered on O, and that
$\sigma_3$ is predominantly of the Pb $6s$-type. However, below we
do not impose any constraints on the MO~LCAO coefficients for
these molecular orbitals, based on this expectation.  Note that
only the orbitals $\sigma_3$ and $\pi_{1,2}$ contribute to the
spin-density of the molecular states under consideration. Thus, we
do not need to know the orbitals $\sigma_{1,2}$, and below we omit
the index for the orbital $\sigma_3$.

Now we specify coefficients of the MO~LCAO expansion for the three
valence orbitals of interest:
\begin{subequations}
\label{b2a}
\begin{eqnarray}
        &&\!\! |\sigma_{\omega} \rangle =
        S_s |6s_{1/2,\omega} \rangle
\nonumber\\
        &&+S_p \Bigl( -2\omega \sqrt{\frac{1}{3}} |6p_{1/2,\omega} \rangle
        +\sqrt{\frac{2}{3}} |6p_{3/2,\omega} \rangle\Bigr)\!,
\\
        &&\!\! |\pi_{i,\omega} \rangle =
        P_i\Bigl(2\omega \sqrt{\frac{2}{3}} |6p_{1/2,\omega} \rangle
        +\sqrt{\frac{1}{3}} |6p_{3/2,\omega} \rangle\Bigr)\!,
\\
        &&\!\! |\pi_{i,\omega'} \rangle =
        P_i |6p_{3/2,\omega'} \rangle,
\end{eqnarray}
\end{subequations}
where $\omega = \pm 1/2$ and $\omega' = \pm 3/2$. The numerical
coefficients are chosen to account for the quantum number
$\lambda$:  $\lambda = 0(1)$ for $\sigma$($\pi$) orbitals.  In
order to calculate $W_d$ we must determine the 4 parameters in
Eqs.~\eref{b2a}. Below we try to constrain these parameters using
experimental information about states \eref{b1b} and~\eref{b1c}.
To simplify the notation, we define the ground state of the
molecule as vacuum. Then each of the excited states in
Eqs.~\eref{b1b} and~\eref{b1c} is a two-particle state with one
hole and one electron. We do not use any special notation for the
hole states; instead we simply write the hole orbital in front of
the electron one.  We construct wavefunctions of these states from
the orbitals \eref{b2a}, using at the first stage the
$\Lambda,\Sigma$-coupling scheme classification:
\begin{subequations}
\label{b3}
\begin{eqnarray}
        &&|a(1)[^3\Sigma_1^+]\rangle\; =
\nonumber \\
        &&=\frac{1}{\sqrt{2}}
        \bigl(|\pi_{1,\lambda=-1}\pi_{2,\lambda= 1}\rangle
        +|\pi_{1,\lambda= 1}\pi_{2,\lambda=-1}\rangle\bigr)
        |\uparrow\uparrow\rangle
\nonumber \\
        &&=\; \frac{1}{\sqrt{2}}
        \left(|\pi_{1,-1/2}\pi_{2, 3/2}\rangle
        +|\pi_{1, 3/2}\pi_{2,-1/2}\rangle\right);
\label{b3a} \\
        &&|A(0^+)[^3\Pi_0] \rangle\; =
\nonumber \\
        &&=\frac{1}{\sqrt{2}}
        \left(|\sigma\pi_{2,\lambda= 1}\rangle
        |\downarrow\downarrow\rangle
        -|\sigma\pi_{2,\lambda=-1}\rangle
        |\uparrow\uparrow\rangle\right)
\nonumber \\
        &&=\;\frac{1}{\sqrt{2}}\left(|\sigma_{-1/2}\pi_{2, 1/2}\rangle
        -|\sigma_{1/2}\pi_{2,-1/2}\rangle\right);~\textrm{etc.}
\end{eqnarray}
\end{subequations}
We write each wavefunction in both $\lambda\!-\!\sigma$ and
$\omega\!-\!\omega$ representations; the latter is more convenient
for our purposes.

The rules for calculating hole matrix elements follow from the
fact that the hole in the state $|\omega \rangle$ actually means
the absence of the electron in the state $|-\omega \rangle$.
Thus, the expectation value for an electronic operator $\hat{P}$
over the hole state $|\omega\rangle$ can be written as:
\begin{equation}
\langle\omega |\hat{P}|\omega\rangle_h
\equiv -\langle-\omega |\hat{P}|-\omega\rangle_e
\stackrel{T}{=} \mp
\langle \omega |\hat{P}|\omega\rangle_e, \label{b1d}
\end{equation}
where we applied the time-reversal operation $T$. Thus the final
sign depends on the time-reversal symmetry of $\hat{P}$, with the
minus sign corresponding to a $T$-even electronic operator.  For
example, the HFS interaction is given by the product of the
$T$-odd electronic vector $\hat{\bm{A}}\bm{J}_e$ and the nuclear
spin $\bm{I}$. Thus, for the HFS interaction the plus sign
in~\Eref{b1d} is correct.  A similar argument shows that the SO
constant $\xi$ for a hole has the opposite sign as for an
electron.

{}From Eqs.~\eref{b3}, the first-order SO splitting $\Delta_{AB}$
between states $A(0^+)$ and $B(1)$ is:
\begin{equation}
        \!\!
        \Delta_{AB} = \frac{\xi}{2} \left[ \langle\pi_{2,3/2}|
        {\bm ls}|\pi_{2,3/2}\rangle
        \!-\!
        \langle\pi_{2,1/2}|{\bm ls}|\pi_{2,1/2}\rangle
        \right]
        \!=\! \frac{\xi P_2^2}{2}.
\label{c1}
\end{equation}
Using the experimental value of this splitting \cite{HH79} and the
ionic value for $\xi$ from Table~\ref{tab1}, we estimate $P_2$:
\begin{equation}
        P_2^2=
        \frac{2\Delta_{AB}}{\xi} = \frac{2 \cdot 2420}{9450}
        =0.51.
\label{c2}
\end{equation}
We see that the orbital $\pi_2$ has a large contribution from the
Pb orbital $6p$. The data on energy levels \cite{HH79} shows then
that for all levels with one electron in the $\pi_2$ orbital, the
SO interaction is comparable to the splittings between these
levels. Therefore, there must be significant SO mixing between
such states.

We start with the mixing within configuration $\sigma^2 \pi_1^3
\pi_2$. The mixing angle $\alpha$ between states $a(1)$ and
$C'(1)$ is:
\begin{eqnarray}
        \alpha \approx
        \frac{\langle ^3\Sigma_1^-|H_{\rm so}| ^3\Sigma_1^+
        \rangle}{|\Delta_{aC'}|}
        =\frac{\xi_1+\xi_2}{2|\Delta_{aC'}|},
\label{c3a}\quad
\end{eqnarray}
where $\Delta_{aC'}$ is the energy splitting between $a(1)$ and
$C'(1)$, and $\xi_i \equiv \xi P_i^2$. If we assume that $P_1^2
\ll P_2^2$ (corresponding to the naive ionic model), we can
estimate the value of $\alpha$:
\begin{equation}
        \alpha \approx \frac{\xi_2}{2|\Delta_{aC'}|} \approx 0.3,
\label{c3e}
\end{equation}
and write the new wavefunction in the form:
\begin{eqnarray}
        |a(1)\rangle &=& c_\alpha
        |\pi_{1,3/2} \pi_{2,-1/2}\rangle
        + s_\alpha
        |\pi_{1,-1/2} \pi_{2,3/2}\rangle,
\label{c3f}\\
        c_\alpha &\equiv& \cos\left(\frac{\pi}{4}-\alpha\right),
        \quad s_\alpha \,\equiv\,
        \sin\left(\frac{\pi}{4}-\alpha\right).
\end{eqnarray}

SO interaction also mixes configuration $\sigma^2 \pi_1^3 \pi_2$
with configurations $\sigma \pi_1^3 \pi_2^2$ and $\sigma \pi_1^4
\pi_2$. These mixings can be accounted for by substitution of the
original orbitals $|\pi_{i,1/2} \rangle$ with the perturbed
orbitals
\begin{eqnarray}
        |\tilde{\pi}_{i,1/2} \rangle &=&
        c_i |\pi_{i,1/2} \rangle +
        s_i |\sigma_{1/2} \rangle.
\label{b2b}
\end{eqnarray}
There is no experimental information about levels of the
configuration $\sigma \pi_1^3 \pi_2^2$, so we cannot reliably
estimate the mixing parameter $s_2$. In contrast, both levels with
$\Omega=1$ of the configuration $\sigma \pi_1^4 \pi_2$ are known
[i.e., B(1) and D(1)]. That allows us to write for $s_1$ the
estimate:
\begin{eqnarray}
        s_1 &=& 2.8\,s_\alpha^2 P_1 S_p.
\label{c4}
\end{eqnarray}
These SO mixings then lead to the final form of the wavefunction
of the state $a(1)$:
\begin{equation}
        |a(1) \rangle
        = c_\alpha |\pi_{1, 3/2}\tilde{\pi}_{2,-1/2}\rangle
        + s_\alpha |\tilde{\pi}_{1,-1/2}\pi_{2, 3/2}\rangle.
\label{c7}
\end{equation}


The $G$-factor for the state~\eref{c7} is given by:
\begin{equation}
        G_\parallel =
        \langle a(1) |L_0+2S_0 |a(1) \rangle
        \,=\, 2 - s_\alpha^2 s_1^2 - c_\alpha^2 s_2^2.
\label{c8}
\end{equation}
The measured value $G_\parallel=1.84(3)$ \cite{HMU02} corresponds
to the following equation for mixing parameters:
\begin{eqnarray}
        s_\alpha^2 s_1^2 + c_\alpha^2 s_2^2 &=& 0.16(3).
\label{c9}
\end{eqnarray}
The signs of the parameters $s_{1\!,2}$ should be chosen so that
the contribution of atomic orbital $6p_{1/2}$ to the molecular
orbital $\sigma$ is increased: in this case relativistic
corrections to the binding energy of the $\sigma$ orbital are
positive.

The matrix element of the HFS interaction for the state
$a(1)$~\eref{c7} has the form:
\begin{eqnarray}
  &&\langle a(1)|H_{\rm hfs}|a(1) \rangle =
\label{c10}\\&&
  =c_\alpha^2 \left[
  \langle \pi_{1,3/2}|h_{\rm hfs}|\pi_{1,3/2} \rangle
  -\langle \tilde{\pi}_{2,1/2}|h_{\rm hfs}|\tilde{\pi}_{2,1/2} \rangle
  \right]
\nonumber\\&&
  +\,s_\alpha^2 \left[
  \langle \pi_{2,3/2}|h_{\rm hfs}|\pi_{2,3/2} \rangle
  -\langle \tilde{\pi}_{1,1/2}|h_{\rm hfs}|\tilde{\pi}_{1,1/2} \rangle
  \right].
\nonumber
\end{eqnarray}
We use expressions from Ref.~\cite{KL95} for the one-electron
matrix elements and numbers from \tref{tab1}, combined with the
measurement of the hyperfine constant for the state $a(1)$:
$A_\parallel=-4.1$~GHz \cite{HMU02}, to find another equation
relating the various coefficients of the model:
\begin{eqnarray}
   &&30\! \left(c_\alpha^2 s_2^2 + s_\alpha^2 s_1^2\right) S_s^2
   + 1.8\! \left(c_\alpha^2 s_2^2 + s_\alpha^2 s_1^2\right) S_p^2
\nonumber\\&&
   +\! \left(4.6 s_\alpha^2 c_1^2 - 1.4 c_\alpha^2\right) P_1^2
   +\! \left(4.6 c_\alpha^2 c_2^2 - 1.4 s_\alpha^2\right) P_2^2
\nonumber\\&&
   -\, 4.7 s_\alpha^2 c_1 s_1 P_1 S_p
   - 4.7 c_\alpha^2 c_2 s_2 P_2 S_p
   \,=\, 4.1\,.
\label{c12}
\end{eqnarray}
(Note that the formulae of Ref.~\cite{KL95} are strictly
applicable only for orbitals and states with $\omega = \Omega =
1/2$.  Eq. \eref{c12} takes into account simple modifications of
these formulae for the present situation.)

We now have five equations, namely: \eref{c2}, \eref{c3e},
\eref{c4}, \eref{c9}, and~\eref{c12} on seven parameters of the
wavefunction \eref{c7}. That leaves us with two independent
parameters of the model. We introduce two additional constraints,
which account for normalization and the Pauli principle:
\begin{eqnarray}
   && S_s^2 + S_p^2 \le N_0,
   \qquad
   P_1^2 + P_2^2 \le N_0.
\label{c13}
\end{eqnarray}
We choose $N_0=1.2$ here in order to account for inaccuracy of the
Hartree-Fock approximation used to determine the atomic parameters
in \tref{tab1}.


The parameters $\alpha$ and $P_2$ are unambiguously fixed by
Eqs.~\eref{c2} and~\eref{c3e}. We choose $s_1$ and $P_1$ as free
parameters and solve Eqs.~\eref{c4}, \eref{c9}, and~\eref{c12} for
parameters $s_2$, $S_p$ and $S_s$. After that we reject solutions
which do not meet the constraints \eref{c13}. A typical solution
is:
\begin{equation}
\label{p1} \left\{
\begin{array}{lcllcll}
        \alpha&=&0.3;   & P_2 &=&0.714; &  \\
         s_1  &=&0.107; & P_1 &=&0.503; &  \\
         s_2  &=&0.449; & S_p &=&0.349; & S_s =0.812.
\end{array} \right.
\end{equation}
Some of the 
parameters appear relatively well-defined, while
others are not. The variation ranges are:
\begin{equation}
\label{p2} \left\{
\begin{array}{llllllll}
         s_1   &\le&0.2; & 0.4 &\le&s_2 &\le& 0.5  \\
         S_p^2 &\le&0.5; &&& S_s^2&\ge&0.5.
\end{array} \right.
\end{equation}
The parameter $P_1$ appears to be restricted only by the
normalization condition \eref{c13}.

It may be possible to add some restrictions to reduce the ranges
of variation in \Eref{p2}. For example, the relatively large value
of $s_2$ should require a large value of $S_p$. However, such
additional restrictions would add arbitrariness to the model and
may affect its reliability. We use only the minimal set of
constraints to determine the range of possible values of $W_d$.


For the wavefunction~\eref{c7}, there are two contributions to the
EDM parameter $W_d$ from each of the one-electron orbitals with
$|\omega|=1/2$:
\begin{subequations}
\label{d1}
\begin{eqnarray}
        W_d
        &=& -c_\alpha^2 W_d^{\tilde{\pi}_2}
        -s_\alpha^2 W_d^{\tilde{\pi}_1}
\label{d1a}\\
        &=&
        \frac{4 w_{sp}}{\sqrt{3}} S_s
        \Big(\sqrt{2}c_\alpha^2 c_2 s_2 P_2 - c_\alpha^2 s_2^2 S_p
\nonumber\\        &&+\,
        \sqrt{2}s_\alpha^2 c_1 s_1 P_1 - s_\alpha^2 s_1^2 S_p
        \Big)\!.
\label{d1b}
\end{eqnarray}
\end{subequations}
We find that the first term in \eref{d1b} always dominates the
sum. The second term is not negligible, but the final two terms
contribute $\lesssim 10\%$.  It is important that the leading
contribution to $W_d$ is similar to the first term in \Eref{c12},
which dominates the HFS. This implies that the parameter $W_d$ is
well-constrained even though some of the parameters of the
wavefunction are not. We obtain:
\begin{eqnarray}
        |W_d| &=& 16.6 \pm 3.0\; \textrm{a.u.},
\label{d3}
\end{eqnarray}
where the uncertainty reflects the range of values found within
the model just described.

It is also important to check how $W_d$ depends on the ``fixed''
parameters $\alpha$ and $P_2$, as well as on the input data for
$A_\parallel$ and $G_\parallel$, since our model relating the
MO~LCAO coefficients to these parameters is rather crude.  In
\tref{tab_edm} we solve the model equations for values of these
quantities varying from the best values by $\pm 20\%$. We find
that this variation of the input parameters widens the range for
$W_d$ substantially (to $\pm 5.4$ a.u.), but still does not allow
dramatically smaller values of $W_d$.

\begin{table}[tb]
\caption{Dependence of the EDM constant $W_d$ (in a.u.) on the
parameters of the model.}

\label{tab_edm}

\begin{tabular}{dddddd}
\hline \hline \multicolumn{1}{c}{$A_\parallel$}
&\multicolumn{1}{c}{$G_\parallel$} &\multicolumn{1}{c}{$\alpha$}
&\multicolumn{1}{c}{$P_2^{2^{\vphantom{1}}}$}
&\multicolumn{2}{c}{$|W_d|$}   \\
\multicolumn{1}{c}{(GHz)} &&&&\multicolumn{1}{c}{max}
&\multicolumn{1}{c}{min}\\
\hline
 -4.1  &  1.84~ & ~0.30 &  0.51~ & 19.6~  & 13.7~ \\
 -4.1  &  1.81  &  0.30 &  0.51  & 19.1  & 12.0 \\
 -4.1  &  1.87  &  0.30 &  0.51  & 20.3  & 15.4 \\
 -4.1  &  1.84  &  0.24 &  0.51  & 19.8  & 13.2 \\
 -4.1  &  1.84  &  0.36 &  0.51  & 20.3  & 15.5 \\
 -4.1  &  1.84  &  0.30 &  0.41  & 18.2  & 12.1 \\
 -4.1  &  1.84  &  0.30 &  0.61  & 20.3  & 15.1 \\
 -3.3  &  1.84  &  0.30 &  0.51  & 17.0  & 11.2 \\
 -4.9  &  1.84  &  0.30 &  0.51  & 22.0  & 16.4 \\
\hline \hline
\end{tabular}
\end{table}

It is known from previous calculations of $W_d$ for other diatomic
molecules, that correlation corrections tend to decrease the
result by 10--20\% from the Hartree-Fock level. Therefore, we
state our final result as a conservative lower limit on $W_d$:
\begin{eqnarray}
        |W_d| &\ge& 10\; \textrm{a.u.}
        = 12\cdot 10^{24} \frac{\textrm{Hz}}{e\,\textrm{cm}}.
\label{d4}
\end{eqnarray}

This lower bound is several times larger than earlier, naive
estimates which did not consider the effect of SO mixing on the
(nominally) $\pi$-type orbitals of the a(1) state \cite{KL95}. Our
model shows significant similarity between the orbital
$\tilde{\pi}_{2,1/2}$ in PbO and the single valence orbital in the
ground state of the free radical PbF.  It is thus natural that our
bound is close to the value calculated for PbF~\cite{KFD87}.
(Coincidentally, our bound is also similar to the calculated value
for YbF \cite{Koz97,Par98,QSG98,MKT98a}.) However, we stress that
this first semiempirical estimate of the effective field in PbO
has very limited accuracy.  Thus, more elaborate calculations of
the $a(1)$ state are highly desirable.

MK thanks the University of New South Wales
for hospitality and acknowledges support from RFBR, grant
No 02-02-16387.  DD was supported by NSF Grant PHY9987846, a NIST
Precision Measurement Grant, Research Corporation, the David and
Lucile Packard Foundation, and the Alfred P. Sloan Foundation.


\end{document}